\documentclass[aip, jcp, citeautoscript, amsmath, amssymb, reprint,
]{revtex4-2}
\usepackage[title]{appendix}
\usepackage{graphicx, tikz, orcidlink}
\usepackage{epstopdf}
\usepackage{ascmac}
\usepackage{ulem}
\usepackage{cancel}
\usepackage{here}
\usepackage{physics}
\usepackage{bm}
\usepackage{comment}
\usepackage{booktabs}
\usepackage{multirow}

\usepackage{amsthm,color}
\newtheoremstyle{query}%
{}{}
{\color{red}}
{}
{\sffamily\bfseries}{:}{12pt}
{}
\theoremstyle{query}
\newtheorem{aq}{Author Query/Comment}

\newcommand{\baq}{\begin{aq}}
\newcommand{\eaq}{\end{aq}}

\begin{document}
\title{Isotope Effects in 2D correlation infrared Spectra of Water: HEOM Analysis of Molecular Dynamics–Based Machine Learning Models}
\date{Last updated: \today}

\author{Kwanghee Park\orcidlink{0009-0005-5223-8234}}
\affiliation{Department of Chemistry, Graduate School of Science,
Kyoto University, Kyoto 606-8502, Japan}
\author{Ryotaro Hoshino\orcidlink{0009-0004-4208-326X}}
\affiliation{Department of Chemistry, Graduate School of Science,
Kyoto University, Kyoto 606-8502, Japan}
\author{Yoshitaka Tanimura\orcidlink{0000-0002-7913-054X}}
\email[Author to whom correspondence should be addressed: ]{tanimura.yoshitaka.5w@kyoto-u.jp}
\affiliation{Department of Chemistry, Graduate School of Science,
Kyoto University, Kyoto 606-8502, Japan}

\begin{abstract}
We model, simulate, and analyze the intramolecular modes of liquid H$_2$O and D$_2$O to elucidate how energy excitation, relaxation, and vibrational dephasing interplay through anharmonic mode-mode coupling. Our analysis employs two-dimensional (2D) correlation spectra, a representative observable in nonlinear infrared vibrational spectroscopy. Accurate reproduction of these 2D spectral profiles requires not only a precise dynamical description of intramolecular vibrations but also an appropriate treatment of thermal environmental effects arising from strong interactions with surrounding molecules, which act as thermal baths. Capturing the essential features of the 2D spectra further demands a non-Markovian, non-perturbative, and nonlinear description of the interactions between intramolecular modes and their baths.
To this end, we adopt a hierarchical equations of motion (HEOM) framework to compute the 2D spectra. By comparing the resulting spectra of H$_2$O and D$_2$O, we explore the underlying mechanisms governing their complex energy and phase relaxation dynamics.
\end{abstract}

\maketitle

\section{Introduction}
\label{sec:intro}

Water molecules possess a simple molecular structure, yet the collective
dynamics mediated by hydrogen bonding are remarkably complex, underpinning a
wide range of chemical reactions and biological processes.\cite{OCSACR1999}
Isotope substitution provides a fundamental means of probing these intricate
dynamics. Replacing H$_2$O with D$_2$O modifies zero-point energies and
vibrational couplings, leading to measurable changes in spectral line shapes,
coherence lifetimes, and reaction kinetics. Such isotope-dependent variations
have been exploited to elucidate hydrogen-bonding networks,\cite{Paesani2019}
clarify tunneling contributions,\cite{Cukierman2006} and refine mechanistic
models of aqueous reactions.\cite{Kohen2010} 
Ultrafast spectroscopic techniques,
including X-ray free-electron laser experiments, offer direct access to the
quantum properties of hydrogen-bonded liquids by probing femtosecond dynamics
and isotope-dependent vibrational coupling.\cite{Guillemin2023} 
In particular, two‑dimensional vibrational spectroscopy (2DVS)\cite{ElsaesserH2O, ElsaesserCPL2005, ElsaesserDwaynePNAS2008, Tokmakoff2002, Tokmakoff2003H2O, TokmakoffH2O, Tokmakoff2015, Tokmakoff2016H2O, D2OTokmakoff2016, Tokmakoff2022, ACR2009Pshenichnikov, Hamm2011, Kuroda_BendPCCP2014} has proven to be a powerful approach for investigating isotope effects, as it provides direct experimental evidence for vibrational coherence dynamics and their lifetimes within hydrogen‑bonded water networks.\cite{Yagasaki_ARPC64, YagasakiSaitoJCP20082DIR, Skinner2DIR2007HODPNAS, JansenSkiner2010, JansenChoShinji2DVPerspe2019}

Recent theoretical studies have further shown that sparse hydroxyl solutes
(e.g., HOD in liquid D$_2$O) serve as clean spectroscopic probes of the local
hydrogen-bonding environment, owing to the OH stretching vibration's relative
isolation from other modes.\cite{HynesH2,Skinner2002HOD1,Skinner2002HOD2,
Skinner2003HOD3,Skinner2003HOD4,SkinnerStochs2003,Skinner2004HOD,Skinner2005HOD,TokmakoffGeisslerforHB2003,JansenPshenichnikov2009,VothTokmakoff_St-BendJCP2017}
Molecular dynamics (MD) simulations combined with empirical mappings between
local electric fields and OH stretching frequencies have yielded valuable
insights into IR, Raman, and 2D IR spectra.\cite{Imoto_JCP135,
ImotXanteasSaitoJCP2013H2O,Imotobend-lib2015,PaesaniJCTC2013H2O,
PaesaniJCTC2014BH2O,PaesaniJCP2014qMD,PaesaniJCTC2014H2O,PaesaniJCTC2014H2O}
However, discrepancies remain, particularly in spectral linewidths and echo
peak-shift dynamics, highlighting the limitations of empirical
frequency-field correlations and classical approximations in capturing the
quantum nature of environmental interactions and spectral diffusion.
\cite{Paesan2018H2OCMD,JianLiu2018H2OMP,Althorpe2019CMD}

Given the inherent difficulty of simulating nonlinear spectroscopy within a quantum MD framework, we have recently developed a suite of computational tools designed to overcome this challenge. These tools employ machine-learning (ML) techniques to parameterize a multimodal anharmonic Brownian (MAB) model directly from MD trajectories (\texttt{sbml4md}). \cite{UT20JCTC,PJT25JCP1,PUT26JCP1}  The resulting model provides a physically transparent platform for analyzing the roles of anharmonicity, energy relaxation, and vibrational dephasing as they interplay through anharmonic mode-mode coupling among intramolecular modes.

Capturing the essential features of the 2D spectra further demands a non-Markovian, non-perturbative, and nonlinear description of the interactions between intramolecular modes and their baths in both classical and quantum regimes.\cite{IT08CP}  Therefore, in parallel with this research, our group developed a computational platform capable of rigorously treating (i) any two quantum inter‑ and intramolecular modes (\texttt{DHEOM-MLWS}),\cite{TT23JCP1,TT23JCP2} (ii) any three classical inter‑ and intramolecular modes (\texttt{CHFPE‑2DVS}),\cite{HT25JCP1,HT25JCP2} and (iii) three quantum intramolecular modes (\texttt{HEOM‑2DVS}),\cite{HT26JCP1} thereby enabling the numerical computation of 2DVS.  By combining these tools, we can model various solutions, including water, and obtain spectra such as 2D IR-Raman and 2D IR as needed. In this paper, we utilize this framework to model, simulate, and analyze the 2D correlation spectroscopy of H$_2$O and D$_2$O, thereby clarifying the underlying mechanisms governing their complex energy and phase relaxation dynamics.

This paper is organized as follows. Section \ref{sec:EOM for MAB} introduces our strategy, which combines MD and ML to parameterize the MAB model and compute 2D IR spectra via HEOM: $\mathrm{MD} \rightarrow \mathrm{ML}\ (\texttt{sbml4md}) \rightarrow \mathrm{MAB\ model} \rightarrow \mathrm{2D\ IR}\ (\texttt{CHFPE‑2DVS} / \texttt{HEOM‑2DVS}).$
Section \ref{sec:Calculation} explains simulation details.  Then Sec. \ref{Results_discussion} presents the evaluated parameter values of the MAB model for H$_2$O and D$_2$O, along with the linear absorption and 2D correlation IR spectra calculated from them. Section \ref{sec:conclusion} is devoted to the summary.

\section{MD Simulation, ML‑Based MAB Modeling, and 2D IR via HEOM}
\label{sec:EOM for MAB}

\begin{figure*}[t]
  \centering
  \includegraphics[width=\textwidth]{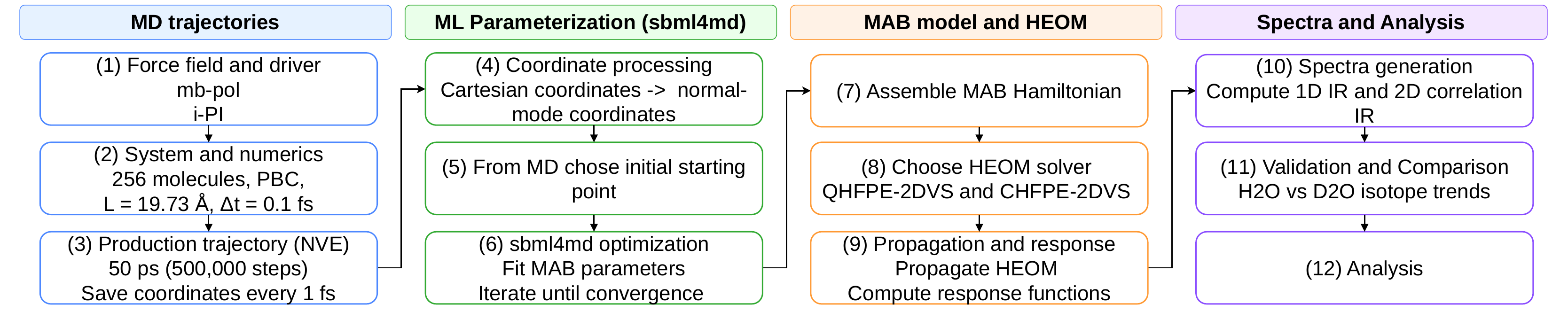}
  \caption{Overall workflow used in this study: MD trajectory generation, ML-based parameterization of the MAB model (\texttt{sbml4md}), assembly of the S-B Hamiltonian, \texttt{CHFPE/HEOM} propagation, and construction/analysis of 1D and 2D correlation IR spectra.}
  \label{fgr:workflow}
\end{figure*}

Here, we briefly describe our complementary strategy for constructing data‑driven models from atomic trajectories and for computing 2DVS by solving them within the HEOM framework. Figure \ref{fgr:workflow} summarizes the workflow used in this study.
For clear illustration, we restrict the present discussion to the intramolecular modes of water, although the approach is in principle applicable to any solvent–solute system or to reaction centers within biomolecules.

\subsection{MAB model}
At the core of our approach is the MAB model.  In this paper, we consider the three primary intramolecular modes of the water molecule: (1) anti-symmetric stretch, ($1'$) symmetric stretch, and (2) bending.  
These modes are described by dimensionless vibrational coordinates $\bm{q}=(q_1, q_{1'}, q_2)$. Each mode is independently coupled to the other optically inactive modes, which constitute a bath system represented by an ensemble of harmonic oscillators. The total Hamiltonian can then be expressed as\cite{UT20JCTC,PJT25JCP1,PUT26JCP1,IIT15JCP,IT16JCP,TT23JCP1,TT23JCP2,HT25JCP1,HT25JCP2,HT26JCP1} 
\begin{eqnarray}
\hat{H}_{tot} &=  \sum_{s} \left( \hat{H}_{A}^{(s)}
+\hat{H}_{I}^{(s)} +\hat{H}_{B}^{(s)} +\hat{H}_{C}^{(s)} ) \right) \nonumber \\
&+  \frac1{2} \sum_{s\ne s'}  \hat{U}_{ss'}\qty(\hat{q}_s, \hat{q}_{s'}),
\label{sec:Total Hamiltonian}
\end{eqnarray}
where
\begin{eqnarray}
\hat{H}_{A}^{(s)}= \frac{\hat{p}_s^{2}}{2m_s} +\hat U_s(\hat{q}_s)
\label{sec:System Hamiltonian}
\end{eqnarray}
is the Hamiltonian for the $s$th mode ($s=1, 1',$ and 2), with mass $m_s$, coordinate ${\hat{q}_s}$, and momentum ${\hat p_s}$; and
\begin{eqnarray}
\hat U_s(\hat{q}_s)=  \frac{1}{2}  m_s \omega_s^2 \hat{q}_s^2 +\frac{1}{3!}g_{s^3}q_{s}^3
\label{sec: Potenentials}
\end{eqnarray}
is the anharmonic potential for the $s$th mode, described by the frequency  $\omega_s$ and cubic anharmonicity $g_{s^3}$. 
The mode-mode coupling between the $s$th and $s'$th modes is given as 
\begin{eqnarray}
\hat{U}_{ss'}(\hat{q}_s, \hat{q}_{s'}) =  g_{s{s'}}\hat{q}_s\hat{q}_{s'} + \frac{1}{6}  \qty(g_{s^2s'}\hat{q}_s^2 \hat{q}_{s'} + g_{s{s'}^2} \hat{q}_s \hat{q}_{s'}^2 ),\nonumber \\
\label{sec: Potential ss'}
\end{eqnarray}
where $g_{s{s'}}$ represents the second-order harmonicity, and $g_{s^2s'}$ and $g_{s{s'}^2}$ represent the third-order anharmonicity. 
In the third-order response function considered below, the contributions from even-order anharmonicity vanish.\cite{OT97JCP1} Therefore, here we retain only the third-order anharmonic terms.

The bath Hamiltonian for the $s$th mode is expressed as\cite{T06JPSJ,T20JCP}
\begin{eqnarray}
\hat{H}_{B}^{(s)}= \sum_{j_s}\qty(\frac{\hat{p}_{j_s}^{2}}{2m_{j_s}}+\frac{m_{j_s}\omega_{j_s}^{2} \hat{x}_{j_s}^2}{2} ) ,
\label{sec: Bath Hamiltonian}
\end{eqnarray}
where the momentum, coordinate, mass, and
frequency of the $j_s$th bath oscillator are given by ${p}_{j_s}$, ${x}_{j_{s}}$, $m_{j_{s}}$ and
$\omega _{{j_s}}$, respectively.
The counter term is expressed as\cite{TW91PRA,OT97PRE}
\begin{eqnarray}
\hat{H}_{C}^{(s)}=  \Lambda^{(s)} \hat{V}_s^2(\hat{ q}_s)
\label{sec: counter Hamiltonian}
\end{eqnarray}
with $ \Lambda^{(s)}  \equiv \sum_{j_s} {\alpha_{j_s}^2 }/{2m_{j_s} \omega _{j_s}^2 }$.
The system-bath (S-B) interaction, defined as
\begin{eqnarray}
  {H}^{(s)}_{\mathrm{I}}&=- V_{s}(\hat {q_s})\sum _{j_s}\alpha _{j_s}{\hat x}_{j_s},
  \label{eq:h_int}
\end{eqnarray}
consists of linear-linear (LL)\cite{TM93JCP,T98CP}
and square-linear (SL) S-B interactions,\cite{TS20JPSJ,KT04JCP,OT97PRE}
$V_{s}({\hat q_s})\equiv V^{(s)}_{\mathrm{LL}}{\hat q_s}+V^{(s)}_{\mathrm{SL}}{\hat q_s^2}/2$, with coupling strengths $V^{(s)}_{\mathrm{LL}}$, $V^{(s)}_{\mathrm{SL}}$,
and $\alpha _{j_s}$. For a vibrational mode with weak anharmonicity, the LL interaction leads to energy relaxation, while the SL interaction results in vibrational dephasing.\cite{T06JPSJ,TS20JPSJ}

The bath property is characterized by the spectral distribution function (SDF), defined as $J_s (\omega) \equiv \sum_{j_s} ({\alpha^2_{j_s}}/{2 m_{j_s} \omega_{j_s}}) \delta (\omega-\omega_{j_s}) $.To employ the HEOM formalism, we consider the Drude SDF, 
\begin{equation}
J_s(\omega) = \frac{m_s \zeta_s}{2\pi} \frac{\gamma_s^2 \omega}{\omega^2 + \gamma_s^2}.
\label{eq:drude}
\end{equation}
The factor of the counter term is now given by
\begin{eqnarray}
\Lambda^{(s)} = \frac{m_s \zeta_s \gamma_s}{2}.
\label{counterterm}
\end{eqnarray}

\subsection{Generation of MD Trajectories}

The parameter variables of the MAB model are evaluated based on atomic trajectories generated by MD simulations. 
Several important considerations are summarized below.\cite{IJT15SD,HT08JCP,HT11JPCB} 

\textbf{System size.}
Because we aim to model fast intermolecular modes that originate from short-range intermolecular interactions, it is not necessary to perform large-scale simulations with many molecules.
Thus, relatively small system sizes are sufficient, and even systems containing 200--300 molecules reproduce the qualitative behavior of liquid water.

\textbf{Thermostat effects.}
Because 2DIR signals are highly sensitive to molecular dynamics, the use of thermostats such as the Nos\'e--Hoover or Langevin schemes can distort the 2D spectral profiles, particularly along the $t_2$ waiting-time axis.
For accurate dynamical behavior, microcanonical (NVE) simulations are therefore preferred.

\textbf{Force-field dependence.}
The 2DIR response strongly depends on the accuracy of the force field.
Differences in anharmonicity or structural fidelity lead to noticeable changes in the nonlinear response, even for nonpolar systems.
Reliable force fields should therefore be validated against experimental structural data.

\textbf{Long-range treatment of induced polarizability.}
The 2DIR profile is sensitive to the functional form of the induced polarizability.
For many molecular systems, the Ewald summation is required to properly account for long-range dipole--dipole interactions, although the magnitude of the effect depends on the molecule.

\textbf{Atomic trajectories for machine learning.}
For liquid water, a training trajectory has a total length of
approximately 50~ps, with Cartesian coordinates of all atoms in all molecules
saved every 10 steps (1~fs) with an integration time step of 0.1~fs.
The required length of the underlying MD trajectory is evaluated
from the linear absorption spectrum obtained directly from the simulation.
Accurately reproducing the bending-mode peak requires a longer trajectory than
that needed to capture the stretching-mode peak.

\textbf{Notes for other molecular systems.}
Although the discussion above focuses on liquid water, similar considerations apply to other molecular systems. The required trajectory length, the sensitivity to thermostatting, and the importance of long-range electrostatics all depend on the characteristic timescales and interaction strengths of the system under study. For systems with slower structural relaxation or stronger intermolecular correlations, substantially longer MD trajectories may be necessary to achieve converged spectral features. Careful validation against experimental observables is therefore essential when extending the MAB parameterization to different molecular environments.

\subsection{ML procedure (\texttt{sbml4md})}
\label{sbml4md}

The dynamical characteristics of the MAB model are defined by several key components: the vibrational potentials of the individual modes, the anharmonic mode–mode couplings, the functional forms of the linear and nonlinear interactions between the modes and the bath, and, critically, the SDF that governs the bath fluctuations.
In our approach, these interaction forms are specified in advance, and their corresponding dynamical parameters are determined through ML optimization. The bath, treated as a continuum of degrees of freedom, is represented by a set of harmonic oscillators---typically numbering more than one thousand---whose coupling constants are optimized to reproduce the MD results, from which the SDF parameters are subsequently inferred. With these elements in place, the algorithm developed by our group (\texttt{sbml4md}) proceeds for the intramolecular vibrations of liquid water as a representative example.

\subsubsection{Intramolecular Modes}

By representing the molecular environment as a continuous function $J(\omega)$ defined by Eq. \eqref{eq:drude}, we have effectively considered a low-temperature heat bath with infinite specific heat, modeled by an infinite number of harmonic oscillators.
In our description, the intramolecular modes (1), (1') and (2) are described in terms of the two O–H bond lengths and the H–O–H bond angle of the $k$th water molecule, defined as
\begin{eqnarray}
  r_1^k= \left| \mathbf{x} _\mathrm{O}^k -  \mathbf{x} _\mathrm{H_1}^k \right|, \label{eq:def_molcoord1}
    \end{eqnarray}
  \begin{eqnarray}
  r_2^k = \left| \mathbf{x} _\mathrm{O}^k -  \mathbf{x} _\mathrm{H_2}^k\right|, \label{eq:def_molcoord2}
  \end{eqnarray}
and
  \begin{eqnarray}
 \theta^k = \arccos \left( \frac{\left(  \mathbf{x} _\mathrm{O}^k -  \mathbf{x}_\mathrm{H_1}^k \right)
  \cdot\left( \mathbf{x} _\mathrm{O}^k -  \mathbf{x}_\mathrm{H_2}^k\right)}{r_1r_2}\right),
  \label{eq:def_molcoord3}
\end{eqnarray}
where $ \mathbf{x}_\mathrm{O}$, $ \mathbf{x}_{\mathrm{H}_1}$, and  $ \mathbf{x}_{\mathrm{H}_2}$are the positions of the oxygen, the 1st, and 2nd hydrogen atoms, respectively, describing the intramolecular motion of the $k$th molecule and $r_1$ and $r_2$ are the average length of $r_1^k$ and $r_2^k$

The system coordinates for three intramolecular modes are expressed as
\begin{eqnarray}
&&q_\mathrm{1}^k = \frac{1}{2}\left(r_1^k - r_2^k\right), \label{eq:def_modecoord2}\\
&&q_\mathrm{1'}^k = \frac{1}{2}\left(r_1^k + r_2^k - 2r_0\right), \label{eq:def_modecoord1}\\
&&q_\mathrm{2}^k = \theta^k - \theta_0,
  \label{eq:def_modecoord3}
\end{eqnarray}
where the equilibrium values of $r_0$ and $\theta_0$ are evaluated from the equilibrium distributions as $\left<r\right>$ and $\left<\theta\right>$.
The system and S-B interaction, $H^{(s)}_\mathrm{B}$ and $H^{(s)}_\mathrm{I}$, are described with these system coordinates.
The intramolecular mode-mode interaction, $U_{s,s'}(q_s, q_{s'})$, in the system Hamiltonian can be obtained by rewriting the intramolecular potentials
 appearing in Eqs.\eqref{sec: Potenentials} and \eqref{sec: Potential ss'} in terms of Eqs. \eqref{eq:def_modecoord2}-\eqref{eq:def_modecoord3}, respectively. 
 
 \subsubsection{Bath oscillators}
 
The degrees of freedom associated with the surrounding molecular motions, including the representative intermolecular vibrational mode, are treated as bath oscillators and are described by $H_\mathrm{I}^{(s)}$ and $H_\mathrm{B}^{(s)}$, given in Eqs.~\eqref{sec: Bath Hamiltonian} and \eqref{eq:h_int}, respectively.

Within the ML framework, introducing a bath with infinitely many degrees of freedom is not feasible. Consequently, the bath coupled to the $s$th mode is represented by a system of $N_s$ finite harmonic oscillators. The inverse of the interval $\Delta\omega_{N_s}$ between the eigenenergies of these oscillators determines the required sampling time of the trajectory, with longer times needed for smaller intervals. In practice, $N_s$ is typically chosen between 1000 and 2000.

Under this prescription, the trajectory of the $j_s$th oscillator is given by
\begin{equation}
\tilde{x}_{j_s}(t) = A_{j_s}\sin(\omega_{j_s} t + \phi_{j_s}),
\end{equation}
where $\phi_{j_s}$ is a random phase and $A_{j_s}$ is a parameter learned within the ML 
framework. The amplitudes of the bath oscillators $\{A_{j_s}\}$ are utilized to determine the Drude SDF parameters 
$(\lambda_s, \gamma_s)$, and $\{A_{j_s}\}$ are updated consistently with 
SDF throughout training. Because the SDF is defined through the anti-symmetric function of 
$\tilde{x}_{j_s}(t)$ as 
$\sum_{j_s} \langle \{ \tilde{x}_{j_s}(t'), \tilde{x}_{j_s}(t) \} \rangle$, where 
$\{ \cdot , \cdot \}$ denotes the Poisson bracket,\cite{T06JPSJ} the SDF must be optimized with 
respect to $\langle \{ V_s(t'), V_s(t) \} \rangle$ rather than the operator 
$V_s(\hat q_s) = V^{(s)}_{\mathrm{LL}} \hat q_s + V^{(s)}_{\mathrm{SL}} \hat q_s^{2}/2$. 
This indicates that $V_{\mathrm{SL}}^{(s)}$ cannot be determined independently of the 
Drude SDF strength $\zeta$.

Because the overall scale of the SL coupling can be absorbed into the bath strength, 
there is a scale redundancy between the S-B coupling strength and $V_{\mathrm{SL}}^{(s)}$. 
To make this explicit, we factor out the SL coefficient and rewrite the interaction operator as
$\hat V_s =V_{\mathrm{SL}}^{(s)} (\hat q_s^{2}+\bar V_{\mathrm{LL}}^{(s)}\,\hat q_s$ with $\bar V_{\mathrm{LL}}^{(s)} \equiv {V_{\mathrm{LL}}^{(s)}}/{V_{\mathrm{SL}}^{(s)}}.$

We then optimize the coefficients through
\begin{equation}
c_{j_s}^k = \alpha_{j_s} V_{\mathrm{LL}}^{(s)} A_{j_s},
\end{equation}
which allows $\bar V_{\mathrm{LL}}^{(s)}$ to vary separately. 
We subsequently fix the gauge by applying a post-training normalization 
$V_{\mathrm{SL}}^{(s)} \rightarrow 1$ and absorb its magnitude into the bath-strength 
parameter as $\zeta_s \rightarrow \zeta_s \left(V_{\mathrm{SL}}^{(s)}\right)^2$.

In the SL-only calculations reported in this work, we set $V_{\mathrm{LL}}^{(s)}=0$, so that $\bar V_{\mathrm{LL}}^{(s)}=0$ and the coupling operator reduces to $\hat V_s=\hat q_s^{2}/2$ after normalization.

The bath parameters and S-B couplings are then encoded in the latent variables
\begin{equation}
\mathbf{z}_k = \left(
  \{c^k_{j_1}\},
  \{c^k_{j_{1'}}\},
  \{c^k_{j_2}\}
\right),
\label{eq:def_zk}
\end{equation}
where $\{c^k_{j_s}\}$ denotes the set of bath-coupling parameters.

 \subsubsection{ML optimization}

The trajectory of each molecule at time step $i$ is then propagated using the MAB model:
\begin{eqnarray}
&&\left(\tilde{\mathbf{q}}^k(t_0 + i\Delta t), \tilde{\mathbf{p}}^k(t_0 + i\Delta t)\right) 
 =  \hat{L}(\Delta t; \mathbf{z}_k, \Sigma) \nonumber \\ 
 &&~~~~~~~~~~~\times
 (\tilde{\mathbf{q}}^k(t_0 + (i-1)\Delta t), \tilde{\mathbf{p}}^k(t_0 + (i-1)\Delta t)),\nonumber \\ 
\end{eqnarray}
where $\left(\tilde{\mathbf{q}}^k(t), \tilde{\mathbf{p}}^k(t)\right)$ denote the coordinates and momenta of the $k$th molecule.
The operator $\hat{L}(\Delta t; \mathbf{z}_k, \Sigma)$ is the Liouvillian corresponding to 
Eqs.~\eqref{sec:Total Hamiltonian}--\eqref{eq:h_int} with the discretized heat bath, and 
$\Sigma$ denotes the set of system and bath parameters.

For each molecule, phase-space trajectories 
\((\mathbf{q}^k(t), \mathbf{p}^k(t))\) from MD are compared with 
MAB-predicted trajectories 
\((\bar{\mathbf{q}}^k(t), \bar{\mathbf{p}}^k(t))\).
Parameters in Eqs.\eqref{sec: Potenentials} and \eqref{sec: Potential ss'} and the SDF 
Eq.~\eqref{eq:drude} are optimized 
to reproduce the reference MD data.
We define the loss function as the Mean Squared Error (MSE) between the predicted and actual MD trajectories for the $s$th mode:
\begin{eqnarray}
\text{MSE}_{q_s} &\equiv \frac{1}{N} \sum_{i=1}^{N} \left[ \tilde{q}_s^k(t_i) -  q_s^k(t_i) \right]^2.
\end{eqnarray}

Minimization of the above loss functions corresponds to the optimization of the learning model parameters. These include the anharmonicity of the potential energy surfaces, anharmonic mode-mode couplings, coupling strengths for LL and SL interactions, and the SDF parameters associated with each vibrational mode.

To optimize the MAB model parameters, we employ a generative ML scheme:
\begin{enumerate}
  \item Read generated MD trajectories of water molecules from the input file.
  \item Simulate corresponding trajectories with the MAB model under trial parameters.
  \item Compute a loss function measuring deviation from MD references.
  \item Backpropagate the loss to iteratively refine parameters.
\end{enumerate}

\subsection{CHFPE-2DVS}
\label{HEOM_Drude}
Once the MAB variables are obtained, the spectrum is calculated using the HEOM framework. For Eqs.~\eqref{sec:Total Hamiltonian}–\eqref{counterterm} with the Drude SDF [Eq.~\eqref{eq:drude}], \texttt{CHFPE-2DVS}\cite{HT25JCP1} and \texttt{QHFPE-2DVS}\cite{HT26JCP1} 
have been established to treat three vibrational modes to simulate 2D correlation IR spectra.

In the \texttt{CHFPE-2DVS}, the terms associated with the Matsubara frequencies vanish, and the equations are expressed as follows\cite{IT16JCP,HT25JCP1,HT25JCP2}
\begin{eqnarray}
\label{eq:clHEOM}
  \frac{ \partial{W^{(\bm{n})}(\bm{q}, \bm{p}; t)}}{\partial t}& = 
	 (\hat{L}(\bm{q}, \bm{p}) -\sum_{s} n_{s} \gamma_{s}) W^{(\bm{n})}(\bm{q}, \bm{p}; t) \nonumber \\
	& +\sum_{s}\hat{\Phi}_{s} W^{(\bm{n}+\bm{e}_{s})}(\bm{q}, \bm{p}; t)\nonumber \\
	&+\sum_{s}\hat{\Theta}_{s} W^{(\bm{n}-\bm{e}_{s})}(\bm{q}, \bm{p}; t),
\end{eqnarray}
where $W^{(\bm{n})}(\bm{q}, \bm{p}; t)$ is the Wigner distribution function (WDF) and the hierarchical elements are denoted as $\bm{n} = (n_1, n_{1'}, n_2)$, where each $n_s$ is a non-negative integer representing the $s$th mode, and $\bm{e}_s$ is the unit vector in the $s$th mode. 
Note that $W^{(\bm{n})}(\bm{q}, \bm{p}; t)$ has physical meaning only when $\bm{n} = (0, 0, 0)$; for all other values of $\bm{n}$, it serves as an auxiliary WDF that accounts for non-perturbative and non-Markovian S-B interactions.\cite{T06JPSJ,T20JCP}

The classical Liouvillian $\hat{L}$ corresponding to the system Hamiltonian $H_{\rm sys}(\bm{q}, \bm{p}) \equiv \sum_{s} {H}_{A}^{(s)} + \sum_{s<s'} U_{ss'}\qty(\hat{q}_s, \hat{q}_{s'})$ is defined as
\begin{align}
\label{eq:CL_liouville}
\hat{L}(\bm{q}, \bm{p})  W(\bm{q}, \bm{p}) &\equiv \{ H_{\rm sys}(\bm{q}, \bm{p}) , W(\bm{q}, \bm{p})  \}_{\mathrm{PB}}, 
\end{align}
where $\{ \cdot, \cdot \}_{\mathrm{PB}}$ denotes the Poisson bracket, given by
\begin{align}
\{ A, B \}_{\mathrm{PB}} \equiv \sum_s \left( \frac{\partial A}{\partial q_s} \frac{\partial B}{\partial p_s} - \frac{\partial A}{\partial p_s} \frac{\partial B}{\partial q_s} \right)
\end{align}
for any functions $A$ and $B$.

The operators $\hat{\Phi}_s$ and $\hat{\Theta}_s$ describe the energy exchange between the $s$th mode and its corresponding bath, and are defined as follows:\cite{TS20JPSJ,KT04JCP}
\begin{align}
\label{eq:Phi}
\hat{\Phi}_s = \frac{\partial V_s(q_s)}{\partial q_s} \frac{\partial}{\partial p_s}, 
\end{align}
and
\begin{align}
\label{eq:Theta}
\hat{\Theta}_s = \frac{m_s \zeta_s \gamma_s}{\beta} \frac{\partial V_s(q_s)}{\partial q_s} \frac{\partial}{\partial p_s}
+ \zeta_s \gamma_s p_s \frac{\partial V_s(q_s)}{\partial q_s},
\end{align}
where $\zeta_s$ is the S-B coupling strength, $\gamma_s$ is the inverse correlation time, and $T$ is the temperature.

\subsection{HEOM-2DVS}

The system Hamiltonian can always be expressed in matrix form using the energy eigenstates of $\hat{H}_{A}^{(s)}$, denoted as
$\left| n \right\rangle_s$ with eigenenergy $ \hbar \omega _{n}^s= {}_s\left\langle {n } \right|  \hat{H}_{A}^{(s)} \left| {n' }\right\rangle{}_s$. Then for $\hat H_{S} \equiv \sum_{s} \hat{H}_{A}^{(s)} +\sum_{s<s'} \hat{U}_{ss'}\qty(\hat{q}_s, \hat{q}_{s'})$ we have
\begin{eqnarray}
\hat H_S &=& \hbar  \sum\limits_{s}  \sum\limits_{n} \omega _{n}^s  \left| {n } \right\rangle{}_s {}_s\left\langle {n } \right| 
\nonumber \\
&+& \hbar  \sum_{s<s'} \sum\limits_{n \ne {n'} } \Delta _{n n'\, mm'}^{ss'} 
\left| {m} \right\rangle{}_{s'} \left| {n } \right\rangle{}_s  {}_s\left\langle {n'} \right|  {}_{s'} \left\langle {m'} \right|, 
\end{eqnarray}
where $\hbar \Delta _{nn'\,mm'}^{ss'} =  {}_s\left\langle {n } \right|  {}_{s'} \left\langle {m } \right| \hat{U}_{ss'}(\hat{q}_s, \hat{q}_{s'}) 
\left| {n' } \right\rangle{}_s \left| {m'} \right\rangle{}_{s'}
$.

The system part of the S-B interaction is expressed as 
\begin{equation}
\hat V_s =  \sum\limits_{n \ge {n'} }  V_{n n' }^s \left| {n } \right\rangle{}_s {}_s \left\langle {n' } \right|,
\label{eq:sys_interaction}
\end{equation}
where $V_{n n' }^s \equiv {}_s\left\langle {n } \right| V_{s}({{\hat q}_s})  \left| {n' } \right\rangle{}_s $.

The HEOM obtained by transforming the phase-space representation into the energy-eigenvalue representation differs from the standard HEOM in that it requires explicitly including the counter term
\begin{align}
\label{eq:counter1}
\hat H_C= \hbar \sum_{s}  \sum\limits_{n \ge {n'} } \delta_{n {n' }}^{s} \left| {n } \right\rangle{}_s {}_s\left\langle {n' } \right|
\end{align} 
with
\begin{align}
\label{eq:counter2}
\hbar  \delta_{n {n' }}^{s} = \frac{m_s \zeta_s \gamma_s}{2}  {}_s\left\langle {n } \right| V_{s}^2 ({{\hat q}_s})  \left| {n' } \right\rangle{}_s 
\end{align}
 as  $\hat H_{S}' \equiv \hat H_{S} + \hat H_C$  in the system Hamiltonian.

Here and hereafter, we define the hyperoperators \(\hat{A}^{\times} \hat{B} \equiv \hat{A} \hat{B} - \hat{B} \hat{A}\) and \(\hat{A}^{\circ} \hat{B} \equiv \hat{A} \hat{B} + \hat{B} \hat{A}\), for arbitrary operators \(\hat{A}\) and \(\hat{B}\).
The \texttt{HEOM-2DVS} is then expressed as\cite{PJT25JCP1}
\begin{eqnarray}
\label{eq:HEOM_DB}
\frac{d}{dt} \hat{\rho}_{\{{\bf n}_s\}} &=& -\left[ \frac{i}{\hbar} \hat{H}_S'^{\times} + \sum_s \sum_{k=0}^{K_s} \left( n_k^s \nu_k^s \right) \right] \hat{\rho}_{\{{\bf n}_s\}} \nonumber \\
&& - i \sum_s \sum_{k=0}^{K_s} n_k^s \hat{\Theta}_k^s \hat{\rho}_{\{{\bf n}_s - {\bf e}_s^k\}} \nonumber \\
&& - i \sum_s \sum_{k=0}^{K_s} \hat{V}_s^{\times} \hat{\rho}_{\{{\bf n}_s + {\bf e}_s^k\}}.
\end{eqnarray}

The hierarchy elements are indexed by the set $\{{\bf n}_s\} \equiv ({\bf n}_1, {\bf n}_{2}, {\bf n}_3)$, where each ${\bf n}_s$ is a multi-index defined as ${\bf n}_s = (n_0^s, n_1^s, \cdots, n_{K_s}^s)$ for the three-mode case. All elements $\hat{\rho}_{\{{\bf n}_s\}}(t)$ with any negative index $n_k^s < 0$ are set to zero.

The notation $\{{\bf n}_s \pm {\bf e}_s^k\}$ indicates an increment or decrement of the $k$th component of ${\bf n}_s$, where ${\bf e}_s^k$ is the unit vector corresponding to the $k$th frequency component in the $s$th bath. The operators are defined as follows:

\begin{eqnarray}
\label{Pade1}
\hat{\Theta}_0^{(s)} &&=- i \frac{m_s \zeta_s\gamma_s^2}{2} \hat{V}_s^{\circ}  \nonumber \\
&&+\frac{m_s \zeta_s\gamma_s}{\beta \hbar} 
\left(1+\sum_{k=1}^{K_s} \frac{2\eta_k^s \gamma_{s}^2 }{\gamma_s^2
-{\nu_k^s}^2}\right)\hat{V}_s^{\times},
\end{eqnarray}
and
\begin{eqnarray}
\label{eq:Pade2}
\hat \Theta_{k> 0}^{(s)}=-\frac{m_s \zeta_s \gamma_s^2 }{\beta \hbar }\frac{2 \eta_k^s \nu _k}{ {\gamma_s^2} - \nu _k^2} \hat{V}_s^\times,
\end{eqnarray}
where the parameters \(\eta_k^s\) denotes the Padé-approximated thermal coupling.\cite{hu2010communication}

\subsection{Linear absorption spectrum and 2D correlation IR spectra}
For optical measurements in which the molecular system interacts with a laser field $E(t)$, 
the nonlinear components of the dipole moment are essential for describing 2D spectroscopy. 
Here we assume\cite{IT16JCP,TT23JCP1,TT23JCP2,HT25JCP1,HT25JCP2}
\begin{eqnarray}
\hat{\mu} 
= \sum_s \mu^{s} \hat q_s 
+ \frac{1}{2!} \sum_{s,s'} \mu^{ss'} \hat q_s \hat q_{s'},
\label{NLdip}
\end{eqnarray}
where $\mu_s$ and $\mu_{ss'}$ denote the linear and nonlinear elements, respectively.
In this work, the nonlinear elements $\mu^{ss'}$ were taken from Ref.~\onlinecite{HT25JCP1}.
The linear elements $\mu^{s}$ primarily control the relative band strengths in the calculated 1D/2D signals; therefore, after fixing the MAB dynamical parameters, we refined $\mu^{s}$ by fitting the \texttt{CHFPE}-calculated 1D spectrum to the MD-derived 1D spectrum using an automated search procedure (described in Sec. \ref{LABS}).

For the intramolecular three-mode vibration, the excitation frequency greatly exceeds thermal excitation, so the equilibrium state is taken as the ground state $|{\bf 0}\rangle = |0_1,0_{1'},0_2\rangle$. Applying the dipole operator $\hat{\mu}$ yields the excited state $|{\bf 1}\rangle$, which contains single- and double-excitation components. A second application produces both returns to the ground state and higher excitations, collectively denoted as $|{\bf 2}\rangle$.\cite{HT26JCP1}

In the density operator representation, the first-order response function is\cite{T06JPSJ,T20JCP}
\begin{eqnarray}
R^{(1)}(t_{1}) =\qty(\frac{i}{\hbar })\mathrm{tr}\qty{\hat{\mu}\mathcal{G}(t_{1})\hat{\mu}^{\times }\hat{\rho }^{\mathrm{eq}}},
\label{eq:R1} 
\end{eqnarray}
where $\hat{\mathcal{G}}(t)\equiv \exp[ -(i/\hbar) \hat H_{tot}^{\times} t]$, which represents the Green's function (Liouvillian propagator) of the system in the absence of a laser interaction, and
$\hat{\rho }^{\mathrm{eq}}$ is the equilibrium state.

The linear absorption spectrum is then evaluated as $
I(\omega)=\int^\infty_0 dt R^{(1)}(t) \exp(\i \omega t).$ The third-order response is
\begin{eqnarray}
R^{(3)}(t_3,t_2,t_1) = \qty(\frac{i}{\hbar })^{3}\mathrm{tr}\qty{\hat{\mu}\mathcal{G}(t_{3})\hat{\mu}^{\times}\mathcal{G}(t_{2})\hat{\mu}^{\times }\mathcal{G}(t_{1})\hat{\mu}^{\times }\hat{\rho }^{\mathrm{eq}}},\nonumber \\
  \label{eq:RFTIRI}
\end{eqnarray}
which expands into more than sixteen terms for three energy eigenstates. These include both population states such as $|{\bf 1}\rangle\langle{\bf 1}|$ and coherent states such as $|{\bf 2}\rangle\langle{\bf 1}|$. 

In 2D correlation spectroscopy, coherent contributions, such as $|{\bf 2}\rangle\langle{\bf 1}|$ are excluded\cite{2DCrrJonas2001,2DCrrGe2002,2DCrrTokmakoff2003}. Rephasing parts correspond to population states, while non-rephasing parts involve coherences. For density operators in the energy eigenbasis, separation is achieved by evaluating the specific Liouville pathways associated with excited-state populations and their conjugates, yielding the 2D correlation spectra.  The detail of this procedure is explained in Refs. \onlinecite{T20JCP, HT26JCP1}.

\section{Simulation Details}
\label{sec:Calculation}

\subsection{MD simulation}

Intermolecular interactions were modeled with the \texttt{mb-pol} potential%
\cite{PaesaniJCTC2013H2O,PaesaniJCTC2014BH2O}, a quantum-chemistry--based
many-body framework that reproduces the properties of water across all phases
with near--quantum-chemical accuracy. Training data were generated from
molecular dynamics simulations performed with the \texttt{i-PI} driver%
\cite{litman2024ipi}, which was coupled to an external force engine via the
socket interface under classical conditions.

Simulations were carried out for liquid H$_2$O and D$_2$O systems, each
containing 256 molecules in a cubic box of side length 19.7295~\AA\ under
periodic boundary conditions. The MD trajectories were propagated in the
microcanonical (NVE) ensemble, initialized at 298.15~K. The equations of motion
were integrated with a time step of 0.1~fs for a total of 500{,}000 steps
(50~ps). Center-of-mass translational motion was removed to eliminate global
drift associated with finite-precision integration. Cartesian coordinates were
saved every 10 steps (1~fs), and thermodynamic quantities were recorded at each
step. For D$_2$O, isotopic masses were explicitly assigned at initialization
(O: 15.999~u, H: 1.008~u, D: 2.014~u).

\subsection{ML parameterization}

\begin{table}[!htb]
\caption{\label{tab:ferguson_drude_3bo_bath}
Optimized parameters of the three-mode MAB model with the SL interaction for isotopic water trained from \texttt{mb-pol} potential:   
(1) anti-symmetric stretch, ($1'$) symmetric stretch, and (2) bending for H$_2$O and D$_2$O. 
Here,  $\tilde{\zeta_s}$ denotes the normalized S-B coupling strength, and $\gamma_s$ denotes the inverse correlation time of the bath fluctuations,  $V_{\mathrm{SL}}^{(s)}$ 
denotes the SL interaction ($V_{\mathrm{LL}}^{(s)} = 0$), and $\tilde{g}_{s^3}$ is the cubic anharmonicity for the $s$ vibrational mode, respectively. 
We set the fundamental frequency to $\omega_{0}$ = 4000  cm$^{-1}$. The normalized parameters were defined as $\tilde{\zeta}_s \equiv (\omega_0/\omega_s)^2\zeta_s$,
and  $\tilde{g}_{s^3} \equiv (\omega_s/\omega_0)^3 g_{s^3}$. 
Nonlinear dipole elements are reproduced from J. Chem. Phys. 163, 172501 (2025), with the permission
of AIP Publishing\cite{HT25JCP1} and given by $\tilde{\mu}_{11'}=1.2 \times 10^{-2} $, $\tilde{\mu}_{12}=0 $, and $\tilde{\mu}_{1'2}=0 $ with $\tilde{\mu}_{ss} \equiv (\omega_0/\omega_s)^2 \mu_{ss}$. }
\begin{tabular}{c|ccccccc}
  \hline 
  \hline
  & s & $\omega_s$ (cm$^{-1}$)  & $\gamma_s/\omega_0$ & $\tilde{\zeta}_s$ &  ${V}_{\rm SL}^{(s)}$  & $\tilde{g}_{s^3}$ \\
\hline
\multirow{3}{*}{H$_2$O } &1 &   $3712$ & $1.17\times 10^{-2}$ & $5.65$ & $1$& $-2.70\times 10^{-3}$ \\
&$1'$ & $3574$ & $9.33\times 10^{-3}$ & $4.58$ &  $1$ & $-3.43\times 10^{-3}$ \\
&2 &  $1663$ & $1.24\times 10^{-2}$ & $4.75\times 10^{-1}$  & $1$  &$-5.16\times 10^{-2}$ \\
\hline
\multirow{3}{*}{D$_2$O } &1 & $2706$ & $1.17\times 10^{-2}$ & $8.92$ &$1$&  $-7.10\times 10^{-3}$ \\
&$1'$ & $2626$ & $1.21\times 10^{-2}$ & $6.16$  &$1$ & $-8.08\times 10^{-3}$  \\
&2 & $1213$ & $1.29\times 10^{-2}$ & $3.38\times 10^{-1}$&$1$ & $-1.17\times 10^{-1}$\\
  \hline \hline
\end{tabular}
\end{table}

\begin{table}[!htb]
\caption{\label{tab::LL_bbdb_mode}
Optimized mode–mode coupling parameters of the three-mode MAB model for isotopic water trained from \texttt{mb-pol} potential: (1) anti-symmetric stretch, ($1'$) symmetric stretch, and (2) bending modes. Here, we set $\tilde g_{s's}={g}_{s's}(\omega_0/\omega_s)(\omega_0/\omega_{s'})$, $\tilde g_{s^2 s'}={g}_{s^2 s'}(\omega_0/\omega_s)^2(\omega_0/\omega_{s'})$, and $\tilde g_{s s'^2}={g}_{s s'^2}(\omega_0/\omega_s)(\omega_0/\omega_{s'})^2$.}
\begin{tabular}{c|cccc}
  \hline \hline
& $\mathrm{s-s'}$ & $\tilde{g}_{ss'}$  & $\tilde{g}_{s^2s'}$ & $\tilde{g}_{s{s'}^2}$  \\
 \hline
 \multirow{3}{*}{H$_2$O }& $\mathrm{1-1'}$ & $4.28\times 10^{-1}$ & $1.98\times 10^{-3}$ & $1.37\times 10^{-3}$\\
 & $\mathrm{1-2}$ &   $1.01$ & $7.05\times 10^{-3}$ & $1.45\times 10^{-2}$ \\
 & $\mathrm{1'-2}$ & $1.06$ & $5.88\times 10^{-3}$ & $1.44\times 10^{-2}$ \\
 \hline
 \multirow{3}{*}{D$_2$O }& $\mathrm{1-1'}$ & $7.99\times 10^{-1}$ & $4.98\times 10^{-3}$ & $4.48\times 10^{-3}$  \\
 & $\mathrm{1-2}$ &   $1.90$ & $1.89\times 10^{-2}$ & $4.20\times 10^{-2}$  \\
 & $\mathrm{1'-2}$ & $1.83$ & $1.79\times 10^{-2}$ & $3.51\times 10^{-2}$ \\
\hline \hline
\end{tabular}
\end{table}

Based on the obtained MD trajectories, parameterization of the MAB model via
ML was carried out following the procedure described in
\ref{sbml4md}. In this work, we focus on the case in which
$V_{\mathrm{LL}}$ is set to zero.

The thermal bath associated with each vibrational mode was represented by
$N_s = 1500$ harmonic oscillators whose frequencies were defined as
$\omega_{j_s} = j_s \Delta\omega$ with $\Delta\omega = 1\ \mathrm{ps}^{-1}
\approx 5.31\ \mathrm{cm}^{-1}$. This frequency grid spans up to
$8000\ \mathrm{cm}^{-1}$, thereby covering the overtone region of the OH
stretching mode. The initial phases $\phi_{j_s}^k$ were randomly sampled from
a uniform distribution over $[0,2\pi)$ at each iteration step.

The MAB Hamiltonian [Eq.~\eqref{sec:Total Hamiltonian}] was propagated using a
time step of $0.1$~fs.  The latent variables $\mathbf{z}_k$ were initialized
to zero, and both the system parameters and $\mathbf{z}_k$ were iteratively
optimized. A minibatch size of 25 trajectories was used for each epoch, and
the same set of trajectories was employed throughout the optimization.
Iterations were continued until convergence of the potential parameters was
achieved.
Because MAB parameter optimization is non-convex and sensitive to starting points, we used physically motivated initial values for the system and bath parameters.

Initial fundamental frequencies $\omega_s$ were set from the peak positions of the MD-derived linear absorption spectrum for each intramolecular mode, while the initial bath correlation times ($\gamma_s^{-1}$) and coupling strengths ($\zeta_s$) were chosen to reproduce the qualitative linewidths and relaxation timescales observed in the MD trajectories. In practice, we found that the bending mode is particularly sensitive to the initial bath strength. When the initial value of $\zeta_2$ for the bend is set too large, it induces an unphysically strong S-B coupling, which in turn leads to an artificial splitting of the bending peak in the resulting 1D and 2D spectra.
To avoid this failure, we performed a multi-start search over several initial parameter combinations (especially for $\zeta_2$), and selected the converged solution that (i) achieves stable loss reduction during training and (ii) yields a single-band bending feature with a smaller optimized $\zeta_2$.
This strategy improves robustness without changing the model form, and it is consistent with the physical expectation that the bend should not require anomalously strong S-B coupling to reproduce the MD reference.

Model training was performed using Python 3.9.18 in conjunction with TensorFlow 2.15 and CUDA 12.2. All computations were executed on a system equipped with an Intel Core i9-13900H CPU and an NVIDIA GeForce RTX 4070 GPU. 

The model parameters obtained for the MAB are summarized in Tables \ref{tab:ferguson_drude_3bo_bath} and \ref{tab::LL_bbdb_mode}.

\subsection{CHFPE-2DVS calculations}
\label{sec:spectruman}

The \texttt{CHFPE-2DVS}\cite{HT25JCP2} is time-integrated using the fourth-order
Runge--Kutta method. Equation
\eqref{eq:clHEOM} is discretized using the compact finite-difference
scheme on a non-uniform mesh.

For numerical integration, the hierarchy is truncated according to the
condition $\delta_{\mathrm{tot}} > \Delta_{\bm{n}}/N$, where
$N = \sum_s n_s$ and
\begin{align}
    \Delta_{\bm{n}}
    = \prod_s (n_s)^{-0.05}
      \left( {k_B T m_s \zeta_s} \right)^{n_s}.
\end{align}
By adjusting the number of hierarchical elements, the spectrum can be computed
with the desired accuracy.

The entire integration routine is implemented in CUDA with the CUBLAS library
and executed primarily on the GPU, while host–device memory transfers are minimized. The source
code was run on NVIDIA GEFORCE RTX 3080 (10~GB VRAM) GPU boards hosted on a workstation
equipped with an 11th Gen Intel(R) Core(TM) i9-11900.

\subsection{HEOM-2DVS calculations}
To employ \texttt{HEOM-2DVS}\cite{HT26JCP1}, the vibrational Hamiltonian was discretized using the Lagrange--Hermite mesh method (LHMM; Hermite-DVR) with $(N_x,N_y,N_z)=(50,50,50)$ and mode-dependent coordinate scaling factors $\bm{h}=(h_x,h_y,h_z)=(0.5,0.5,0.75)$.
Here, $\bm{h}$ rescales the dimensionless Gauss--Hermite quadrature nodes $r_{d,j}$ to the physical coordinate grid via $q_{d,j}=h_d\,r_{d,j}$ ($d=x,y,z$), thereby setting the effective coordinate extent of the LHMM grid for each mode.
The DVR product basis was truncated by the cut scheme with $\sum_{d=1}^{3} n_d\le 2$, which yields a 10-dimensional basis prior to final diagonalization.
Cross-mode anharmonic couplings were then included, and the Hamiltonian was diagonalized in this truncated basis.
For the subsequent HEOM propagation, eigenstates with energies $E\le 3.0$ (in the scaled unit defined by $\omega_0=4000~\mathrm{cm^{-1}}$) were retained, resulting in $N_{\mathrm{vib}}=8$ levels for H$_2$O and $N_{\mathrm{vib}}=9$ levels for D$_2$O.
Each vibrational mode was coupled to an independent Drude bath, whose parameters are summarized in Tables~\ref{tab:ferguson_drude_3bo_bath} and \ref{tab::LL_bbdb_mode}.
The bath correlation functions were treated in the pure Drude form without Pad\'e decomposition, i.e., the number of Pad\'e terms was set to $K_s=0$ for all modes.
The HEOM was truncated using the tolerance criterion with $\delta_{\mathrm{tot}}=10^{-15}$.
Time evolution was performed using a fourth-order Runge--Kutta integrator with linear-stability control (RK4LS) with a time step of $dt=0.25$~fs at $T=300$~K. Linear absorption spectra were obtained from trajectories of length 1.2~ps, corresponding to 4800 propagation steps. For 2D correlation spectra, the third-order response functions were evaluated on $t_1$ and $t_3$ windows of 0.6~ps each (2400 steps per time axis), with waiting times $t_2=\{1,50,100,500\}$~fs.

\section{Results and discussions}
\label{Results_discussion}
\subsection{Linear absorption spectra}
\label{LABS}

\begin{figure}[htbp]
  \centering
  \includegraphics[keepaspectratio, scale=0.38]{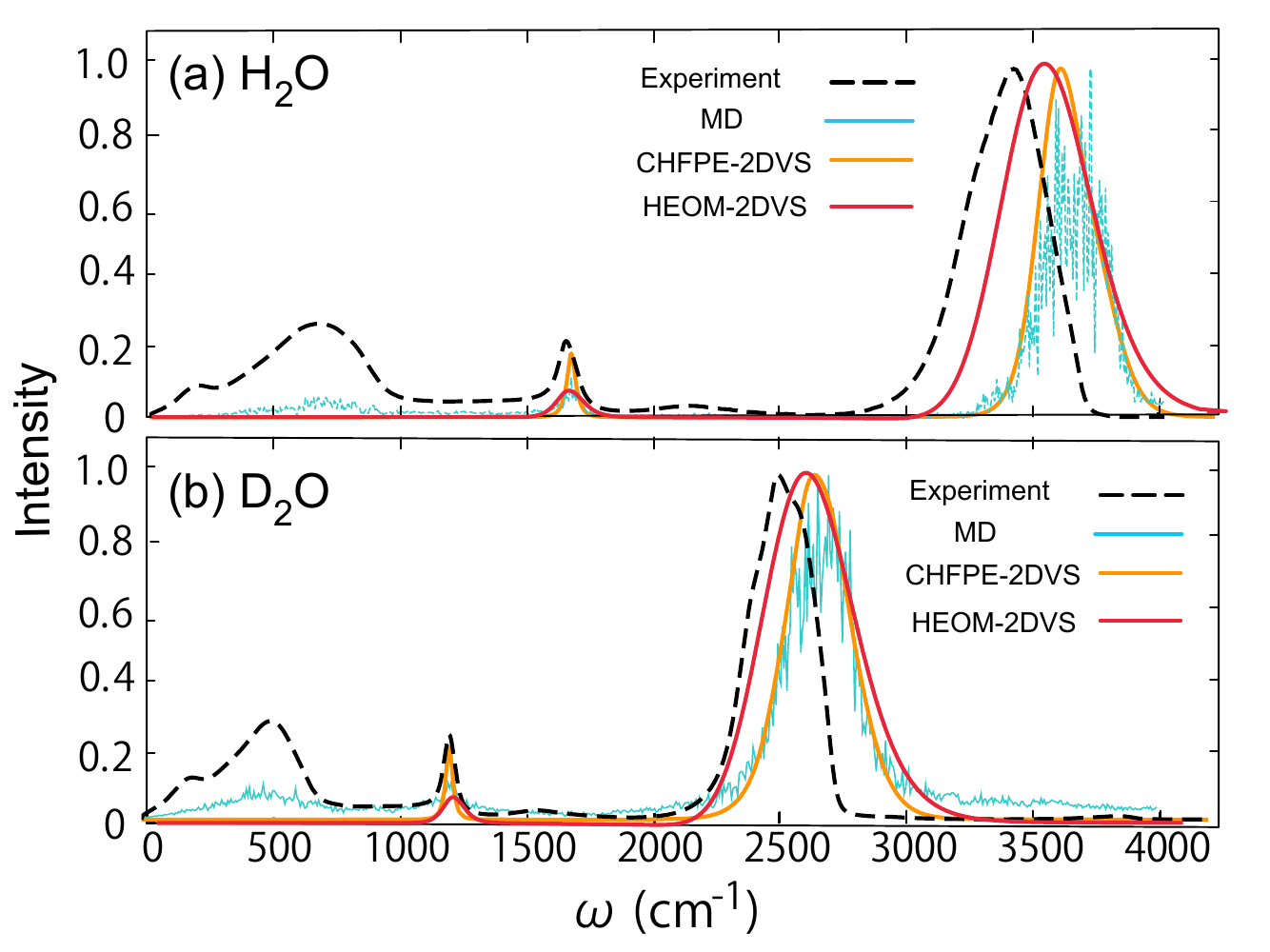}
  \caption{Linear absorption spectra of (a) H$_2$O and (b) D$_2$O calculated using the 
three-mode MAB models with the parameter sets listed in
Tables~\ref{tab:ferguson_drude_3bo_bath} and \ref{tab::LL_bbdb_mode}. 
For comparison, each panel also includes results from MD simulations 
(blue lines) and experimental data (black dashed curves). Each spectrum is normalized to its maximum peak intensity.
The H$_2$O experimental spectrum is reproduced with permission from 
Y. Maréchal, J. Mol. Struct. 1004, 146 (2011). Copyright (2011) Elsevier.\cite{IRexp2011}
The D$_2$O experimental spectrum is reproduced from J. Chem. Phys. 131, 
184505 (2009), with the permission of AIP Publishing.\cite{D2OIR} 
}
  \label{fgr:linear}
\end{figure}

The three‑mode MAB parameters for H$_2$O and D$_2$O extracted from MD trajectories (Tables~\ref{tab:ferguson_drude_3bo_bath} and \ref{tab::LL_bbdb_mode}) were used in the classical 
\texttt{CHFPE‑2DVS} and quantum \texttt{HEOM‑2DVS} calculations. Figure~\ref{fgr:linear} compares the resulting IR spectra with those obtained directly from MD trajectories and with experiment. To obtain the MD spectrum, we used the total dipole moment
$\boldsymbol\mu^{\rm tot}(t)=\boldsymbol\mu^{\rm perm}(t)+\boldsymbol\mu^{\rm ind}(t)$,
evaluated at each sampling step. After removing the time-averaged dipole
$\langle \boldsymbol\mu^{\rm tot}\rangle$, we applied a real-valued FFT to each
Cartesian component. The IR intensity was then computed as
$I(\omega)\propto \omega^2 \sum_{\alpha=x,y,z}|\mu^{\rm tot}_{\alpha}(\omega)|^2$.

Note that to match the relative stretching and bending band strengths, we calibrated the linear dipole coefficients $\{\mu^{s}\}$ in Eq.~\eqref{NLdip} against the MD-derived 1D spectrum using a Bayesian hyperparameter search (TPE). Only the selected dipole parameters were varied while the MAB dynamical parameters were fixed, and the \texttt{CHFPE} 1D spectrum was recomputed in each trial. Because all spectra were normalized before comparison, the optimization targeted relative band strengths and lineshape rather than absolute intensity. The resulting best-fit $\{\mu^{s}\}$ were used in all subsequent 1D and 2D calculations.

The classical spectra obtained with the ML‑parameterized MAB model closely reproduce the MD results, confirming the reliability of the parametrization. For both H$_2$O and D$_2$O, the stretching bands appear blue‑shifted in the classical description because quantum anharmonic effects are absent.\cite{ST11JPCA}

Although the present MAB parameters were derived from classical MD trajectories, the underlying \texttt{mb-pol} force field was originally developed for quantum nuclear dynamics. As a result, quantum \texttt{HEOM‑2DVS} calculations using these parameters yield spectra that more closely resemble those expected from quantum‑MD simulations and show better agreement with experiment than the classical results. A more rigorous validation of this approach will require MAB parameters trained directly on quantum‑MD data.

Overall, aside from slight blue shifts, the \texttt{HEOM‑2DVS} calculations based on the three‑mode MAB model provide a reasonable description of the stretching and bending modes.

As experimentally observed, the linear absorption spectra of H$_2$O and D$_2$O
differ because the heavier deuterium nuclei reduce the vibrational
frequencies. Consequently, the O--D
stretching and bending bands appear at much lower frequencies than the
corresponding O--H modes. Moreover, the lighter H nuclei exhibit larger quantum
fluctuations and faster hydrogen-bond dynamics, which broaden the absorption
lines in H$_2$O, whereas D$_2$O shows narrower bands due to its slower
structural fluctuations. In our numerical calculations, the low-vibration D$_2$O stretch mode exhibits suppressed quantum effects, resulting in both classical and quantum outcomes that more closely match experimental values.

\subsubsection{2D correlation IR spectra (quantum case)}

\begin{figure}[htbp]
  \centering
  \includegraphics[keepaspectratio, scale=0.37]{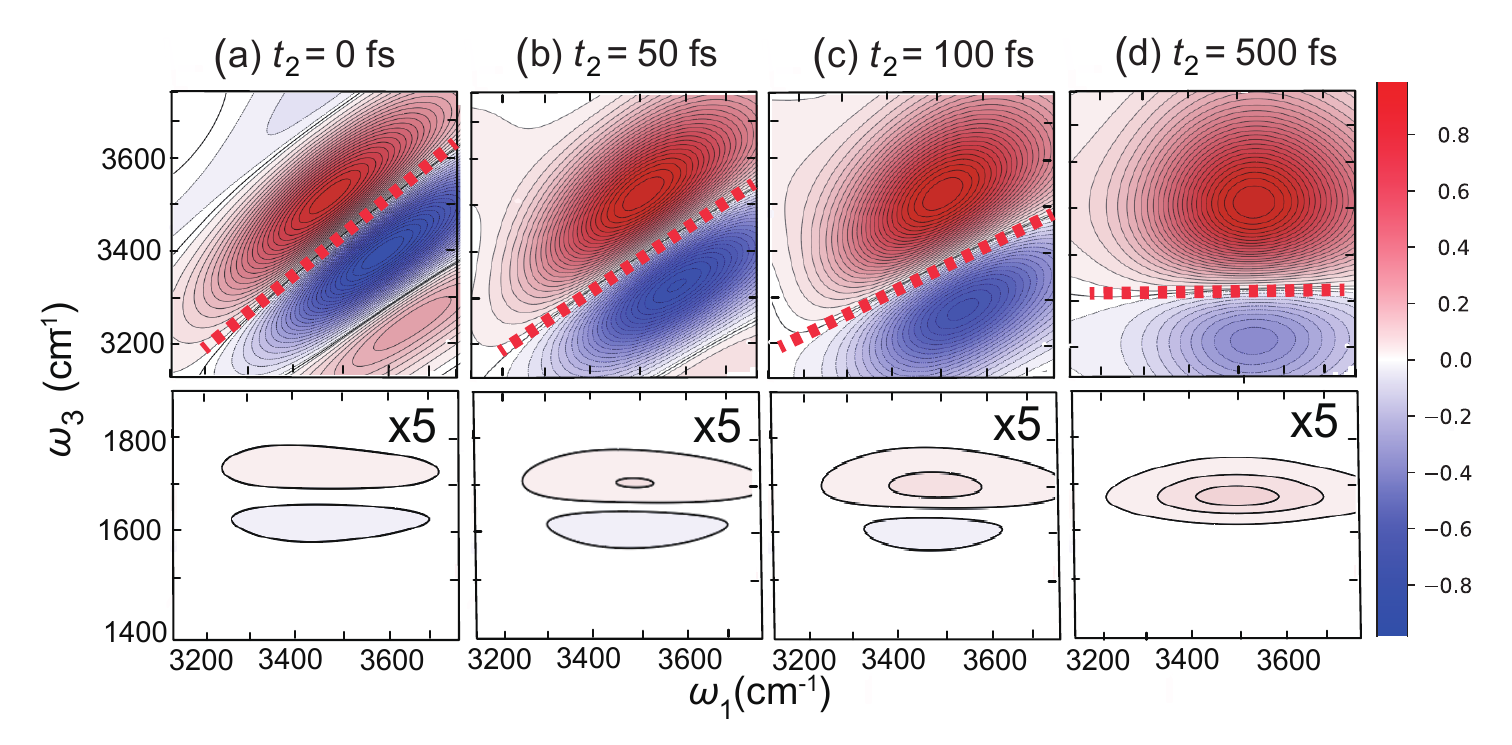}
\caption{The quantum-mechanically computed 2D correlation IR spectra of H$_2$O are shown for the stretching modes (1) and (1$'$) in the upper panel, and for the coupled stretching (1, 1$'$)–bending (2) motions in the lower panels. These spectra were obtained at several $t_2$ periods using the parameter sets listed in Tables~\ref{tab:ferguson_drude_3bo_bath} and \ref{tab::LL_bbdb_mode}. The orientation of the red dashed nodal lines in the upper panel indicates the degree of correlation between the vibrational coherences during the $t_1$ and $t_3$ periods. For clarity, the contour interval in the lower panels has been increased by a factor of five.
}
  \label{fgr:qst-bnH2Ombpol}
\end{figure}

\begin{figure}[htbp]
  \centering
  \includegraphics[keepaspectratio, scale=0.37]{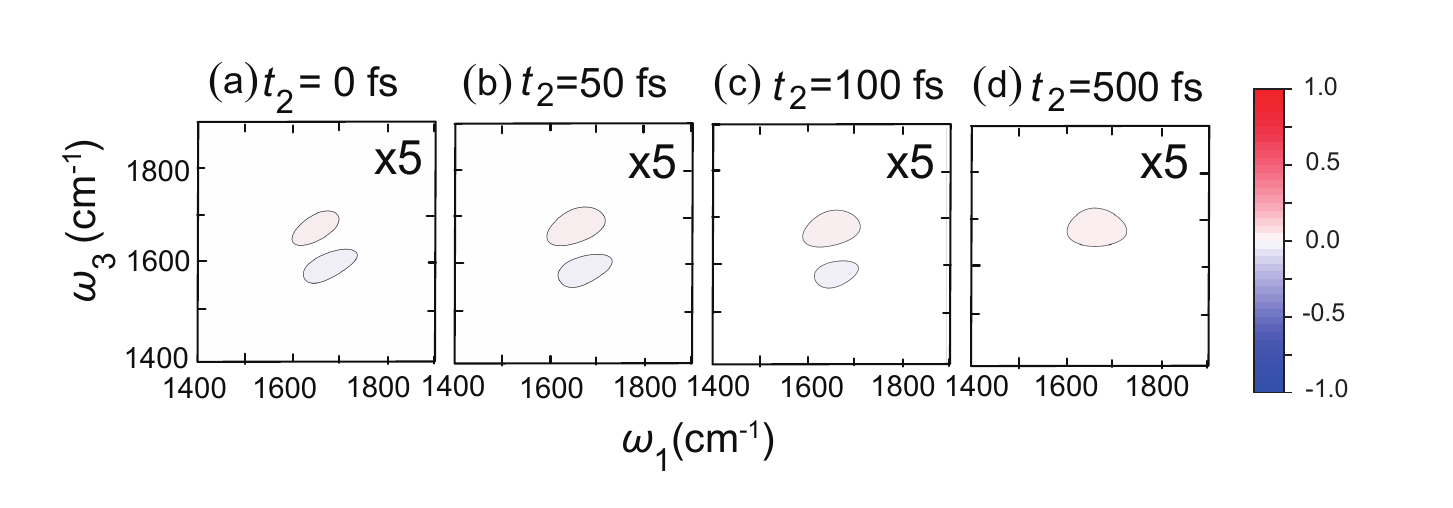}
\caption{The quantum‑mechanically computed 2D correlation IR spectra of H$_2$O for the (2) bending motions were obtained using the parameter sets listed in Tables~\ref{tab:ferguson_drude_3bo_bath} and \ref{tab::LL_bbdb_mode}. The spectral intensities were normalized to the maximum stretching amplitude. For emphasis, the peak intensity of the bending mode was multiplied by a factor of five relative to the intensity shown in Fig.~\ref{fgr:qst-bnH2Ombpol}.
}
 \label{fgr:qbnH2Ombpol}
\end{figure}

\begin{figure}[htbp]
  \centering
  \includegraphics[keepaspectratio, scale=0.35]{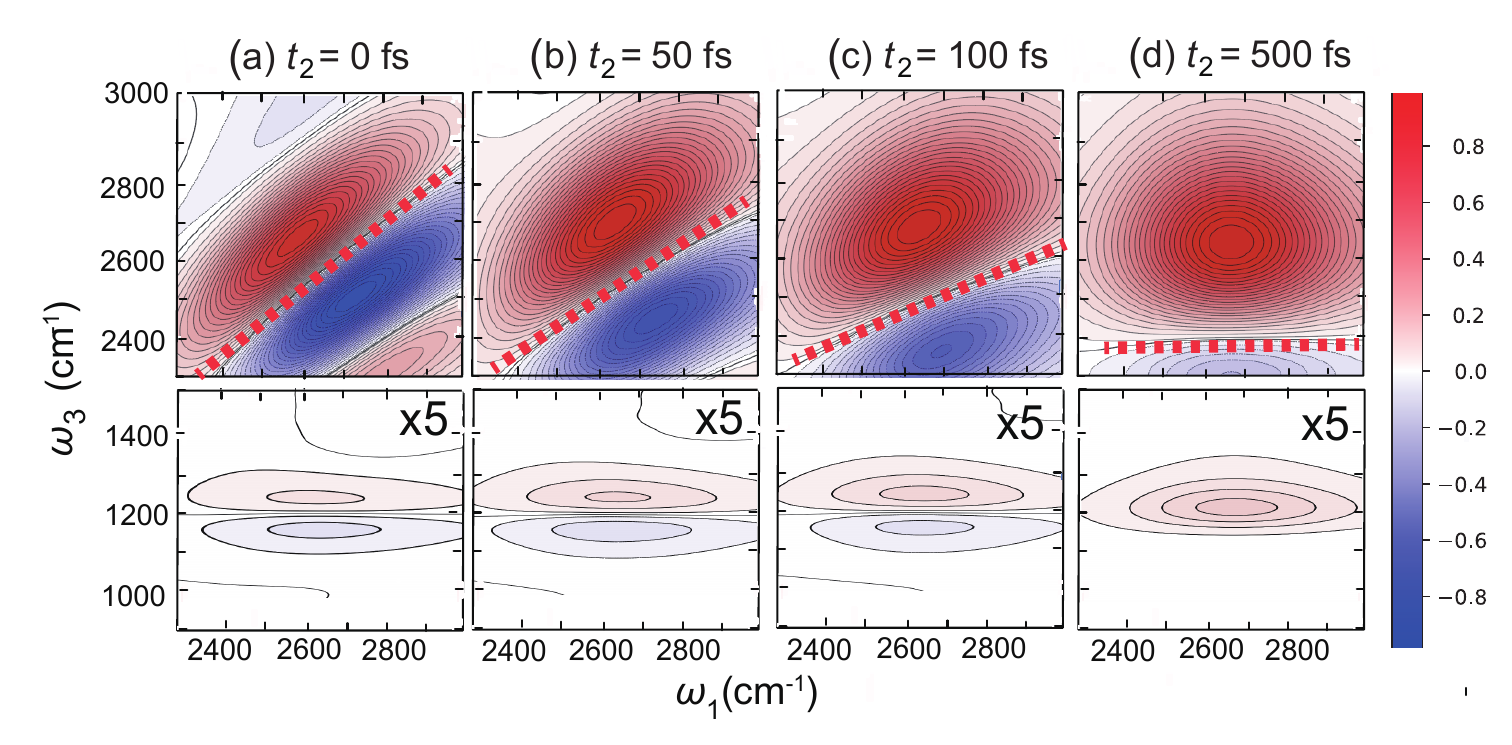}
  \caption{The quantum‑mechanically computed 2D correlation IR spectra of D$_2$O for the stretching modes (1) and (1') [upper panel] and the stretching (1, 1')--bending (2) motions [lower panels] at different $t_2$ periods. The orientation of the red dashed nodal lines in the upper panel indicates the degree of correlation between the vibrational coherences during the $t_1$ and $t_3$ periods. For emphasis, 
 the contour interval in the lower panels was increased by a factor of five.
} 
  \label{fgr:qst-bnD2Ombpol}
\end{figure}

\begin{figure}[htbp]
  \centering
  \includegraphics[keepaspectratio, scale=0.35]{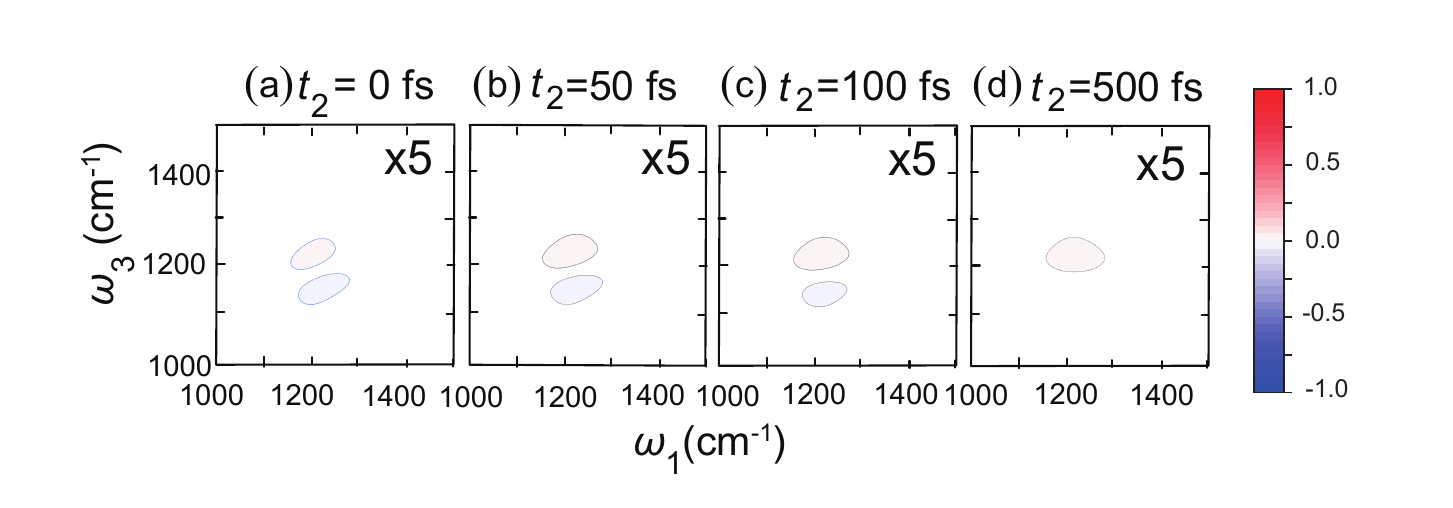}
  \caption{The quantum‑mechanically computed 2D correlation IR spectra of H$_2$O 
for the (2) bending motions using the parameter sets listed in Tables~\ref{tab:ferguson_drude_3bo_bath} and \ref{tab::LL_bbdb_mode}.
For emphasis, the peak intensity of the bending mode was multiplied by a factor of five relative to the intensity shown in Fig. \ref{fgr:qst-bnD2Ombpol}.
}
  \label{fgr:qbnD2Ombpol}
\end{figure}

We now present the quantum-mechanical calculation results for the 2D
correlation IR spectra of H$_2$O and D$_2$O using the three-mode MAB
models whose parameter values are listed in
Tables~\ref{tab:ferguson_drude_3bo_bath} and \ref{tab::LL_bbdb_mode}.
The corresponding classical results are provided in Appendix~\ref{2DIRCL}.

While a number of experimental studies have reported 2D correlated IR spectra involving D$_2$O,\cite{ElsaesserH2O,
ElsaesserCPL2005,ElsaesserDwaynePNAS2008,Tokmakoff2002,Tokmakoff2003H2O,
TokmakoffH2O,Tokmakoff2015,Tokmakoff2016H2O,D2OTokmakoff2016,Tokmakoff2022,ACR2009Pshenichnikov,
Hamm2011,ElsaesserDwaynePNAS2008,Tokmakoff2016H2O,Tokmakoff2022,Kuroda_BendPCCP2014}
 whereas theoretical predictions that simultaneously treat all three intramolecular vibrational modes have remained scarce.\cite{HynesH2,Skinner2002HOD1,Skinner2002HOD2,
Skinner2003HOD3,Skinner2003HOD4,SkinnerStochs2003,Skinner2004HOD,Skinner2005HOD,TokmakoffGeisslerforHB2003,JansenPshenichnikov2009,VothTokmakoff_St-BendJCP2017}

Figures~\ref{fgr:qst-bnH2Ombpol} and \ref{fgr:qbnH2Ombpol} show the 2D correlation spectra of H$_2$O calculated at different $t_2$ periods. 

Regarding the two stretching modes, the vibrational states are defined as $|\mathbf{0}\rangle= |0\rangle_{1}\,|0\rangle_{1'}$, 
$ |{\mathbf{1}}^{\pm}\rangle = 
 (  |1\rangle_{1}\,|0\rangle_{1'} 
  \pm   |0\rangle_{1}\,|1\rangle_{1'}
)/\sqrt{2}$, and $|{\mathbf{2}}^{\pm}\rangle
=  (  |2 \rangle_{1}\,|0\rangle_{1'} 
  \pm   |0\rangle_{1}\,|2\rangle_{1'}
)/\sqrt{2}$.
Then the major red and blue peaks can be assigned to the
$|\mathbf{0}\rangle \rightarrow |{\mathbf{1}}^{+}\rangle \rightarrow |\mathbf{0}\rangle$
and
$|\mathbf{0}\rangle \rightarrow |{\mathbf{1}}^{+}\rangle \rightarrow |{\mathbf{2}}^{+}\rangle$
pathways, respectively.
The small blue and red peaks appear at the onset of the two major peaks
originate from the
$|\mathbf{0}\rangle \rightarrow |{\mathbf{1}}^{-}\rangle \rightarrow |\mathbf{0}\rangle$,
$|\mathbf{0}\rangle \rightarrow |{\mathbf{1}}^{-}\rangle \rightarrow |{\mathbf{2}}^{-}\rangle$, and so forth.\cite{HT26JCP1}
These pathways proceed through coherent dynamics, with the associated peaks
emerging immediately at $t_{2}=0$ and decaying as $t_{2}$ increases.

Such behavior differs from the 2D correlation IR spectrum
calculated using \texttt{CHFPE-2DVS} with the two-mode MAB model and from the
experimental spectrum in the limit of very small $t_2$.
This discrepancy is likely due to poorly optimized dipole elements and inter-stretch mode coupling being
overestimated relative to the value obtained through the ML‑optimized
parametrization.

The orientation of the red dashed nodal lines reflects the degree of noise
correlation (non-Markovian effects) between the vibrational coherences during
$t_1$ and $t_3$.
A direction parallel to the $\omega_1$ axis corresponds to the uncorrelated
limit, whereas alignment along the $\omega_1=\omega_3$ diagonal corresponds to
the fully correlated limit.\cite{2DCrrJonas2001,2DCrrGe2002,2DCrrTokmakoff2003}
The diagonal peaks broaden along the $\omega_1=\omega_3$ direction due to
spectral diffusion driven by the slow noise correlation of the stretching
coordinate. As $t_2$ increases beyond the noise correlation time
($\tau_1 \approx 160$~fs), the correlation between the $t_1$ and $t_3$
coherences weakens, and the nodal line gradually rotates toward the $\omega_1$
axis, indicating a transition from correlated to uncorrelated fluctuations.

For the stretching$\rightarrow$bending cross peaks, the peak shapes appear
uncorrelated because the stretch--bend noise correlation time is much shorter than that of stretch and bend them selves.
As a result, the positive and negative cross peaks are elongated along the
$\omega_1$ direction. The cross‑peak intensity decays at early times because it is dominated by coherence, but it subsequently increases once population transfer sets in.\cite{TT23JCP2,HT26JCP1}

Regarding the bending‑mode results in Fig. \ref{fgr:qbnH2Ombpol}, although the peak intensity is somewhat weaker than in the other calculations, the relaxation profile in the correlation spectrum is consistent with both the \texttt{HEOM-2DVS} results based on the POLI‑2VS model\cite{HT26JCP1} and the experimental observations.\cite{Kuroda_BendPCCP2014}

 In D$_2$O results in Figs. \ref{fgr:qst-bnD2Ombpol} and \ref{fgr:qbnD2Ombpol}, where the excitation frequencies of both the stretching and bending modes are lower than in H$_2$O, differences in the overall spectral profile are expected. 

Excluding differences in spectral profiles arising from variations in the model, 
an examination of the contrast between H$_2$O and D$_2$O shows that, in D$_2$O, 
the narrower spacing between the stretch and bend modes leads to stronger 
stretch–bend coupling. This enhanced coupling facilitates more efficient energy 
transfer and consequently results in a slower decay of the cross peak.

The classical 2D correlation IR results evaluated using \texttt{\texttt{CHFPE-2DVS}} are summarized in Appendix~\ref{2DIRCL}. 
The blue-shifted peak positions, the reduced linewidth broadening, 
and the clearer separation of the two stretching modes---features 
characteristic of classical calculations---appear consistently with 
previous computational studies.\cite{HT25JCP1} 
These spectra are expected to be close to those 
obtained by directly evaluating the nonlinear response function from 
MD trajectories. Although experimentally measured spectra naturally 
agree more closely with the quantum results, the classical 2DIR spectra 
remain useful because they approximate what would be obtained from a 
direct MD-based 2DIR calculation. In this sense, they provide a 
practical reference for extrapolating experimental trends from MD 
simulations.

\section{Conclusion}
\label{sec:conclusion}

In this study, we show that the condensed-phase dynamics of the
intramolecular vibrations in H$_2$O and D$_2$O can be captured with
unprecedented fidelity by integrating the \texttt{sbml4md} approach—which
constructs MAB models directly from MD atomic trajectories—with the
\texttt{HEOM-2DVS} and \texttt{CHFPE-2DVS} methods, which provide a numerically
exact treatment of these models. The resulting linear absorption and
2D correlation IR spectra not only reproduce the essential experimental
signatures but also expose striking isotope-dependent variations in
anharmonicity, intermolecular coupling, and both energy and vibrational
relaxation pathways. These features, largely inaccessible to conventional MD,
underscore the critical importance of explicitly incorporating bath effects to
accurately predict nonlinear spectroscopic responses.

Taken together, these results establish a unified mechanistic framework for
vibrational dynamics in hydrogen-bonded liquids. By enabling systematic control
over physical conditions—such as the character of environmental fluctuations or
the presence of quantum effects—beyond the reach of standard MD simulations,
the present approach offers a powerful route for disentangling the fundamental.

Because the \texttt{sbml4md} strategy employed here is intrinsically based on single MD trajectories, it can nevertheless be directly extended to solute molecules in solution, reaction centers in biomolecular environments, and other heterogeneous systems. By solving the quantum system using MAB model parameters determined from trajectories obtained via quantum MD and propagated with quantum HEOM, it becomes possible to incorporate quantum effects in a physically consistent manner.
These directions represent promising avenues for future investigation.

\section*{Acknowledgments}
Y. T. was supported by JST (Grant No. CREST 1002405000170). K. P. acknowledges a fellowship supported by JST SPRING, the establishment of university fellowships toward the creation of science technology innovation (Grant No.~JPMJSP2110). 
\section*{Author declarations}
\subsection*{Conflict of Interest}
The authors have no conflicts to disclose.

\section*{Data availability}
The data that support the findings of this study are available from the corresponding author upon reasonable request.

\appendix

\section{2D correlation IR spectra (classical case)}
\label{2DIRCL}

\begin{figure}[htbp]
  \centering
  \includegraphics[keepaspectratio, scale=0.35]{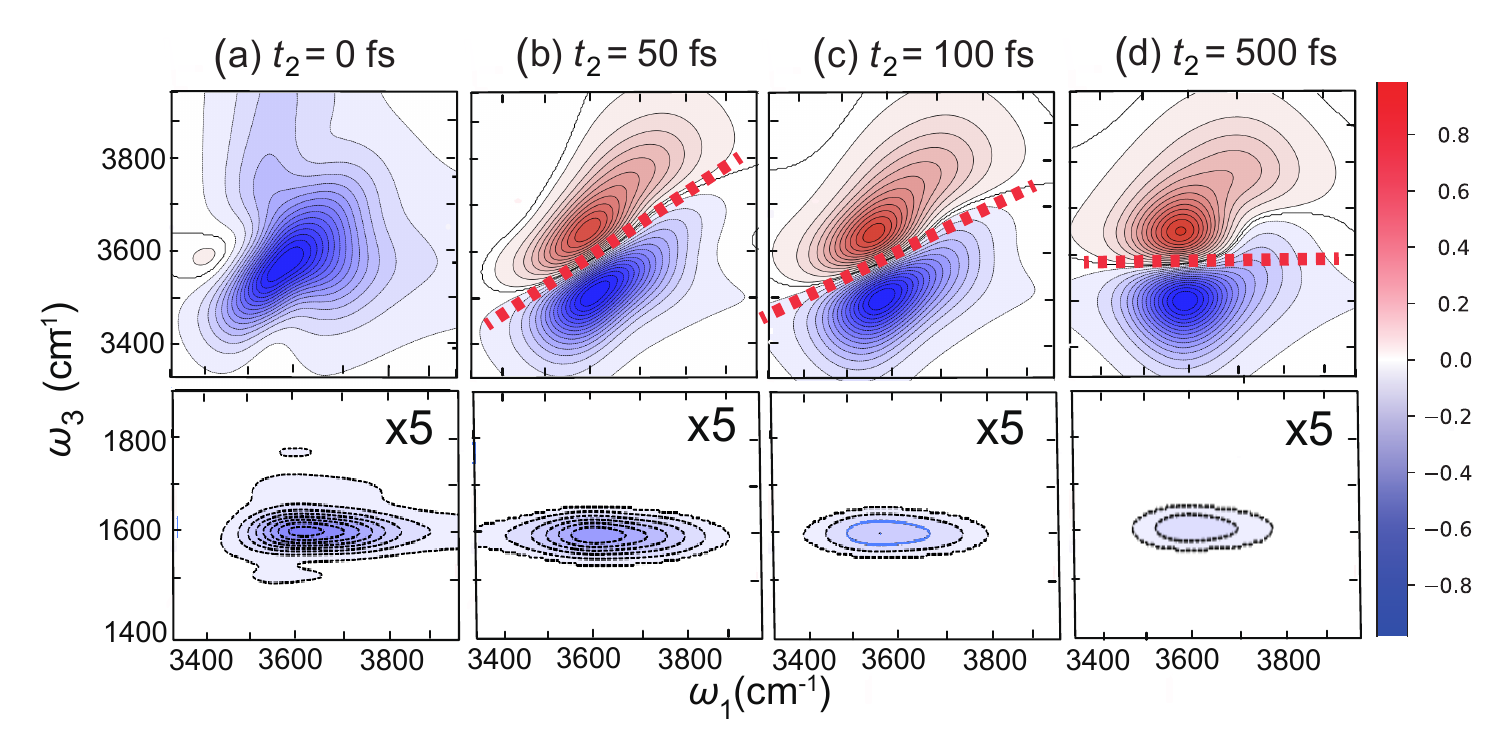}
  \caption{The classical‑mechanically computed 2D correlation IR spectra of H$_2$O for the stretching modes (1) and (1') [upper panel] and the stretching (1, 1')--bending (2) motions [lower panels] at different $t_2$ periods using the parameter sets listed in Tables~\ref{tab:ferguson_drude_3bo_bath} and \ref{tab::LL_bbdb_mode}. The spectral intensities were normalized to the maximum stretching amplitude. For improved visibility, the contour interval in the lower panels---where the peak intensities are weaker---was increased by a factor of five.
}
  \label{fgr:st-bnH2Ombpol}
\end{figure}

\begin{figure}[htbp]
  \centering
  \includegraphics[keepaspectratio, scale=0.37]{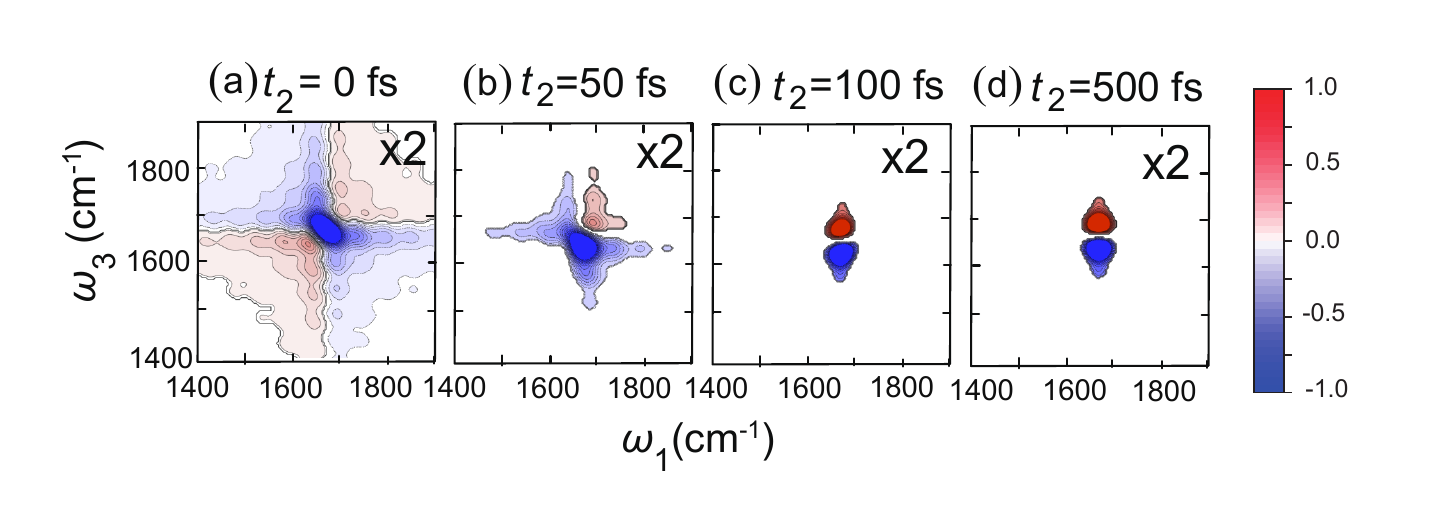}
\caption{The classical‑mechanically calculated 2D correlation IR spectra of H$_2$O for the (2) bending motions using the parameter sets listed in Tables~\ref{tab:ferguson_drude_3bo_bath} and \ref{tab::LL_bbdb_mode}. The spectral intensities were normalized to the maximum stretching amplitude. For emphasis, the peak intensity of the bending mode was multiplied by a factor of two relative to the intensity shown in Fig.~\ref{fgr:st-bnH2Ombpol}.
}
 \label{fgr:bnH2Ombpol}
\end{figure}

\begin{figure}[htbp]
  \centering
  \includegraphics[keepaspectratio, scale=0.35]{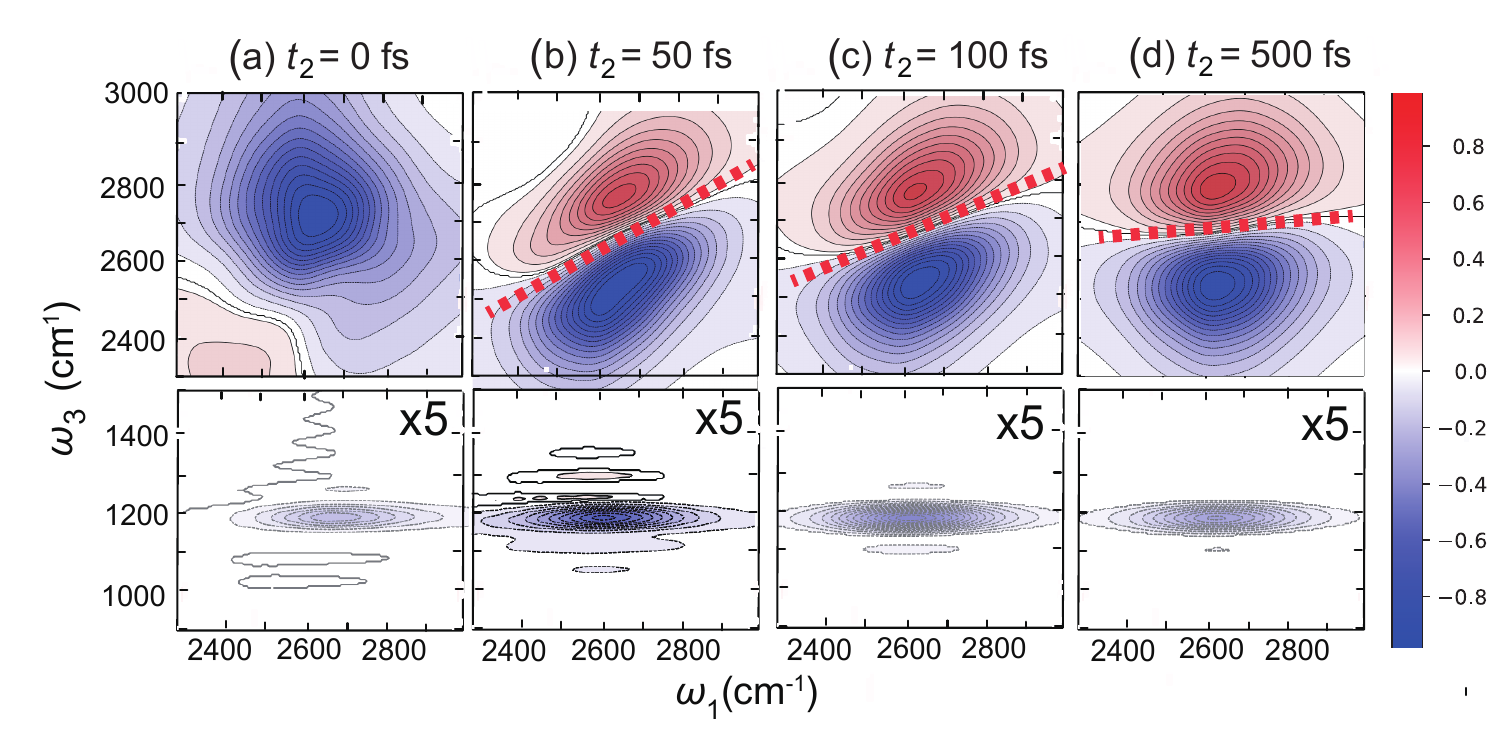}
  \caption{The classical‑mechanically computed 2D correlation IR spectra of D$_2$O for the stretching modes (1) and (1') [upper panel] and the stretching (1, 1')--bending (2) motions [lower panels] at different $t_2$ periods. For emphasis, 
 the contour interval in the lower panels was increased by a factor of three.
} 
  \label{fgr:st-bnD2Ombpol}
\end{figure}

\begin{figure}[htbp]
  \centering
  \includegraphics[keepaspectratio, scale=0.35]{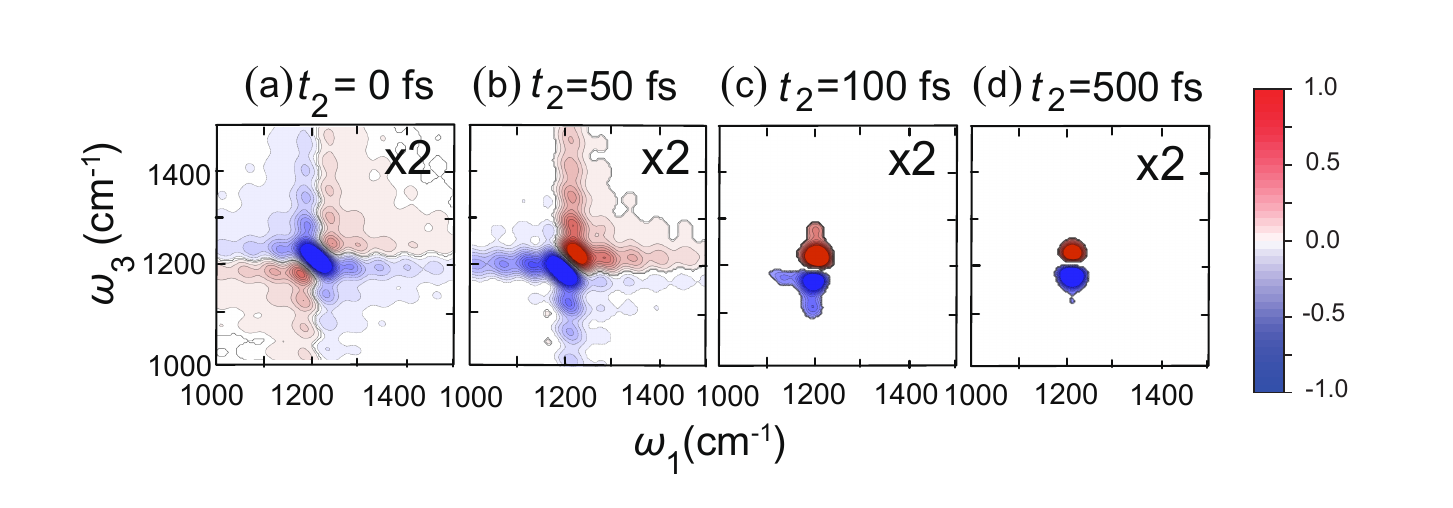}
  \caption{The classical‑mechanically computed 2D correlation IR spectra of H$_2$O  for the (2) bending motions using the parameter sets listed in Tables~\ref{tab:ferguson_drude_3bo_bath} and \ref{tab::LL_bbdb_mode}. The spectral intensities were normalized to the maximum stretching amplitude. For improved visibility, the contour interval in the lower panels---where the peak intensities are weaker---was increased by a factor of two.
}
  \label{fgr:bnD2Ombpol}
\end{figure}

Here, we present the classical mechanical 2D correlation IR spectra of
H$_2$O (Figs.~\ref{fgr:st-bnH2Ombpol} and \ref{fgr:bnH2Ombpol}) and D$_2$O
(Figs.~\ref{fgr:st-bnD2Ombpol} and \ref{fgr:bnD2Ombpol}) obtained using the classical
\texttt{CHFPE-2DVS} method. These calculations employ the three mode MAB
models, with parameter sets listed in
Tables~\ref{tab:ferguson_drude_3bo_bath} and \ref{tab::LL_bbdb_mode}.

The blue shifted peak positions, reduced linewidth broadening, and clearer
separation of the two stretching modes, which are characteristic features of
classical treatments, are consistent with previous computational
studies.\cite{HT25JCP1}

Compared with the quantum mechanical results for H$_2$O, the spectrum at
$t_2 = 0$ differs markedly because coherent transitions are absent in the
classical description. At larger $t_2$ values, the two vibrational modes
become more clearly resolved, as quantum broadening is minimal.

In the classical case, the intensity of the stretch$\rightarrow$bend cross
peak also differs from the quantum result. Because only population relaxation
contributes, the cross peak decays monotonically.
The Bend peaks do not elongate over time because classical dynamics does not
include the effects of coherence.

For low frequency D$_2$O, the results indicate that quantum effects are
small. Thus, even the classical spectra show no substantial differences from the
quantum ones, aside from the overall blue shift of the peaks and the behavior
at $t_2 = 0$.

\bibliography{tanimura_publist.bib,TT23.bib,HT24.bib,PHT26.bib}

\end{document}